\documentclass{article}
\usepackage[utf8]{inputenc}

\usepackage{notes2bib}
\usepackage{graphicx}
\usepackage{amsmath}
\usepackage{url}
\usepackage{amsfonts}
\usepackage{amsthm}
\usepackage[justification=centering]{caption}
\newtheorem{theorem}{Theorem}[section]

\theoremstyle{definition}

\theoremstyle{remark}

\title{Efficient Algorithms for the Closest Pair Problem and Applications}
\author{Sanguthevar~Rajasekaran,
        Sudipta~Pathak\\
Department of CSE, University of Connecticut, Storrs, CT}

\date{July 2014}

\begin{document}
\maketitle
\begin{abstract}
The closest pair problem (CPP)  is one of the well studied and fundamental problems in computing. Given a set of points in a metric space, the problem is to identify the pair of closest points. Another closely related problem is the fixed radius nearest neighbors problem (FRNNP). Given a set of points and a radius $R$, the problem is, for every input point $p$, to identify all the other input points that are within a distance of $R$ from $p$. A naive deterministic algorithm can solve these problems in quadratic time. CPP as well as FRNNP play a vital role in computational biology, computational finance, share market analysis, weather prediction, entomology, electro cardiograph, N-body simulations, molecular simulations, etc. As a result, any improvements made in solving CPP and FRNNP will have immediate implications for the solution of numerous problems in these domains. We live in an era of big data and processing these data take large amounts of time. Speeding up data processing algorithms is thus much more essential now than ever before. In this paper we present algorithms for CPP and FRNNP that improve (in theory and/or practice) the best-known algorithms reported in the literature for CPP and FRNNP. These algorithms also improve the best-known algorithms for related applications including time series motif mining and the two locus problem in Genome Wide Association Studies (GWAS).
\end{abstract}

\section{Introduction}
\label{intro}
The closest pair problem (CPP) has a rich history and has been extensively studied. On an input set of $n$ points, the problem is to identify the closest pair of points. A straight forward algorithm for CPP takes quadratic (in $n$) time. Most of the algorithms proposed in the literature are concerned with the Euclidean space. In the rest of this paper, unless otherwise mentioned, we imply the Euclidean space. In his seminal paper, Rabin proposed a randomized algorithm with an expected run time of $O(n)$ \cite{RAB76} (where the expectation is in the space of all possible outcomes of coin flips made in the algorithm). Rabin's algorithm used the floor function as a basic operation. In 1979, Fortune and Hopcroft presented a deterministic algorithm with a run time of $O(n\log\log n)$ assuming that the floor operation takes $O(1)$ time  \cite{FH79}. Both of these algorithms assume a constant-dimensional space (and the run times have an exponential dependency on the dimension). The algorithm of Preparata and Shamos for points in 2D is deterministic and runs in $O(n\log n)$ time \cite{PS86}. Yao has proven a lower bound of $\Omega(n\log n)$ on the algebraic decision tree model (for any dimension) \cite{YO91}. This lower bound holds under the assumption that the floor function is not allowed.
The algorithm of Khuller and Matias is also randomized and has an expected linear run time utilizing the floor operation \cite{SY95}. There are two steps in the algorithm. In the first step, the distance between the closest pair of points is estimated within a factor of 3. In the second step, the neighborhood of each point $p$ is explored to identify those points that are within a distance of $e$ from $p$, where $e$ is the estimate that step 1 comes up with. Using this neighborhood information, the correct pair is identified.


One of the major issues with the above algorithms is the fact that their run times are exponentially dependent on the dimension.   For example, the expected run time of \cite{SY95}'s algorithm is $O(3^dn)$ on $n$ points from a $d$-dimensional space. So, even for a moderate value of $d$, the algorithm may be very slow in practice. There are numerous applications for which the dimension is very large. In the following sections, we consider two such application domains.


Time series motif mining (TSMM) is a crucial problem that can be thought of as CPP in a large dimensional space. In one version of the TSMM problem, we are given a sequence $S$ of real numbers and an integer $\ell$. The goal is to identify two subsequences of $S$ of length $\ell$ each that are the most similar to each other (from among all pairs of subsequences of length $\ell$ each). These most similar subsequences are referred to as {\em time series motifs}. Let $C$ be a collection of all the $\ell$-mers of $S$. (An $\ell$-mer is nothing but a contiguous subsequence of $S$ of length $\ell$). Clearly, the $\ell$-mers in $C$ can be thought of as points in $\Re^\ell$. As a result, the TSMM problem is the same as CPP in $\Re^\ell$. Any of the above mentioned algorithms can thus be used to solve the TSMM problem. A typical value for $\ell$ of practical interest is several hundreds (or more). For these values of $\ell$, the above algorithms (\cite{RAB76},\cite{FH79},\cite{PS86},\cite{SY95}) will take an unacceptable amount of time (because of the exponential dependence on the dimension). Designing an efficient practical and exact algorithm for the TSMM problem remains an ongoing challenge.

Mueen, et al. have presented an elegant exact algorithm called MK for TSMM \cite{AEQSB09}. MK improves the performance of the brute-force algorithm with a novel application of the triangular inequality. MK is currently the best-performing algorithm in practice for TSMM.
A number of probabilistic as well as approximate algorithms are also known for solving this problem (see e.g., \cite{PMPS08,BES03,TCM07,JJMY08,DCIT07,SG04,YKK05}). For instance, the algorithm of \cite{BES03} exploits algorithms proposed for finding $(\ell,d)$-motifs from biological data. The idea here is to partition the time series data into frames of certain width. Followed by this, the mean value in each frame is computed. This mean is quantized into four intervals and as a result, the original time series data is converted into a string of characters from an alphabet of size 4. Finally, any $(\ell,d)$-motif finding algorithm is applied on the transformed string to identify the time series motifs.

An application of great interest in bioinformatics is Genome Wide Association Study (GWAS). A lot of effort has been spent to identify mappings between phenotypical traits and genomic data. Due to the advent of next generation high throughput sequencing technologies, nowadays it is possible to study the genomic structure of individuals in detail. Single Nucleotide Polymorphisms (SNPs) are  positions in the genome where nucleotides vary among individuals \cite{T07}. SNPs in the human genome are considered to be responsible for different phenotypical traits. In GWAS,  two different problems have been focused on. In single locus association study, researchers try to find out the association between phenotypical traits and individual SNPs. In two locus association study, the goal is to figure out the association between pairs of SNPs and phenotypical traits. A major task in this study is that of identifying the most correlated pair of SNPs. Two locus associations are also known as gene-gene interactions.
Such interactions are believed to be major factors responsible for many complex phenotypical traits \cite{SDNRCNHD07,CDDBDADLMDCJE05,JDLCMPMLYJBEJSTS98,H09,NMDPCGA99,JCSEBSAKLDPJW04}.

Given that the number of SNPs found in humans is $10^5$ to $10^7$, a brute force way of scanning through every possible pair of SNPs to identify the most correlated pair is not feasible in practice. A number of algorithms for the two locus problem can be found in the literature. For instance, genetic algorithms are used in \cite{NYH01} and \cite{YXKR10}. The algorithms proposed in  \cite{XSFW10} and \cite{XFW08} take $O(n^2m+nm^2)$ time, where $n$ is the number of SNPs and $m$ is the number of subjects. An algorithm with an expected run time of $O(mn^{1+\epsilon})$, where $0<\epsilon<1$ is a constant, has been presented in \cite{PBK11}. This algorithm exploits an algorithm known for the  Light Bulb Problem \cite{RSJ89} and Locality Sensitive Hashing (LSH) \cite{M02}.

In this paper we present efficient algorithms for CPP. Our algorithms improve the results reported in several papers including \cite{AEQSB09}, \cite{BES03}, and  \cite{PBK11}. Our main contributions can be summarized as follows:
\begin{enumerate}
\item We improve the CPP algorithm of \cite{AEQSB09} by introducing a novel idea. Specifically, \cite{AEQSB09}'s algorithm is based on two basic ideas. We contribute a third idea that results in an improvement of the run time for exact TSMM by a factor of around 1.5. We also show how to extend our algorithm to solve the fixed radius nearest neighbors problem (FRNNP).
\item We present an algorithm for CPP when the domain of interest has strings of characters (from a finite alphabet) and the metric is Hamming distance. It turns out that MK does not perform well for the case of Hamming distance, and to be fair, we note that the authors of MK do not claim it might. A comparison of our algorithm with MK reveals that our algorithm outperforms MK (by a factor of around 200). Our algorithm can also be used in approximate TSMM. Specifically, instead of using $(\ell,d)$-motif search algorithms, our algorithm can be used in \cite{BES03}. In this case, the run time of the algorithm in \cite{BES03} will improve significantly, since exact algorithms for solving the $(\ell,d)$-motif search problem take time that is exponential in $\ell$ and $d$. Our algorithm is a modified version of the light bulb algorithm of \cite{RSJ89}.
\item The light bulb algorithm of \cite{RSJ89} finds the most correlated pair of bulbs. The light bulb problem can be thought of as CPP in the space of {\bf binary} strings with Hamming distance as the metric. We extend this algorithm when the strings are from an arbitrary (finite) alphabet. More importantly, we present an algorithm for finding the {\bf least} correlated pair of strings (from an arbitrary alphabet). The algorithm of \cite{PBK11} also solves this problem utilizing Locality Sensitive Hashing (LSH) \cite{M02}. Our algorithm does not use LSH. Instead, it uses a novel deterministic mapping function that we have come up with.
\item Using the above algorithm for finding the least correlated pair, we present a novel algorithm for solving the two locus GWAS problem. An experimental comparison reveals that our algorithm is four times faster than the algorithm of \cite{PBK11}. We note here that the authors of \cite{PBK11} use Pearson's correlation coefficient to measure the similarity between a pair of SNPs, whereas we use the complement of the Hamming distance as the measure of similarity.
\end{enumerate}


The rest of this paper is organized as follows. In Section~\ref{sec2} we provide some notations. In Section~\ref{sec3} we present our improved algorithm for CPP (called MPR) and compare it with MK. In this section we also provide an analysis of MK and experimentally compare the performances of MK and MPR. Section~\ref{sec4} deals with the case of character strings and Hamming distance. Specifically, we show how to modify the light bulb algorithm of \cite{RSJ89} to get an algorithm for finding the most correlated pair of strings from an arbitrary finite alphabet. We compare this algorithm with MK experimentally. In Section~\ref{sec5} we present an algorithm for finding the least correlated pair of strings. This algorithm is based on a novel mapping function that we have come up with. Section~\ref{sec6} is devoted to the problem of two locus association in GWAS. In particular, we present a novel algorithm for this problem and compare our algorithm with that of \cite{PBK11}. Some concluding remarks are given in Section~\ref{sec7}.

\section{Notations and Definitions}\label{sec2}
Let $T=a_1,a_2,\ldots,a_n$ be a sequence of real numbers (or characters from a finite alphabet). An $\ell$-mer of $T$ is nothing but a subsequence of $T$ of $\ell$ contiguous elements of $T$. The $\ell$-mers of $T$ are $T_i=a_i,a_{i+1},\ldots,a_{i+\ell-1}$, for $1\leq i\leq (n-\ell+1)$.

If the elements of $T$ are real numbers, then the Euclidean distance between $T_i$ and $T_j$, denoted as $d(T_i,T_j)$, is $\sqrt{\sum_{k=0}^{\ell-1}(a_{i+k}-a_{j+k})^2}$.

If the elements of $T$ are characters from an alphabet $\Sigma$, then the Hamming distance between $T_i$ and $T_j$, denoted as $d(T_i,T_j)$, is $\sum_{k=0}^{\ell-1}\delta(a_{i+k},a_{j+k})$ where $\delta(a,b)=1$ if $a\neq b$ and $\delta(a,b)=0$ if $a=b$ (for any $a,b\in \Sigma$). A sequence of characters can be thought of as a string of characters, since we can obtain a string from the sequence by concatenating the characters. Thus we'll use the terms `a sequence of characters' and `a string of characters' interchangeably.

Let $A=a_1,a_2,\ldots,a_n$ and $B=b_1,b_2,\ldots,b_n$ be two sequences of characters. Also, let the Hamming distance between $A$ and $B$ be $d$. Then, by the {\em number of matches} between $A$ and $B$ we mean $n-d$. Also, the {\em correlation} between $A$ and $B$ is defined to be $\frac{n-d}{n}$.

\section{Time Series Motif Mining Algorithm}\label{sec3}
The input for this problem are a sequence $T=a_1,a_2,\ldots,a_n$ and an integer $\ell$. The goal is to find two $\ell$-mers of $T$  that are the closest to each other (from among all the pairs of $\ell$-mers of $T$). A general version of this problem is one where the input consists of $n$ points from $\Re^\ell$ and we want to identify the two closest points.
A straight forward algorithm will compute the distance between every pair of $\ell$-mers and output the pair with the least distance. Since we can compute the distance between two $\ell$-mers in $O(\ell)$ time, this simple algorithm for TSMM will run in a total of $O(n^2\ell)$ time.

\subsection{MK Algorithm}
The MK algorithm of \cite{AEQSB09} speeds up the brute force method by pruning off a large number of pairs that cannot possibly be the closest. There are two main ideas used in MK.  The first idea in the algorithm is to speedup the computation of distances. Let $x=x_1,x_2,\ldots,x_\ell$ and $y=y_1,y_2,\ldots,y_\ell$ be any two $\ell$-mers. To compute the distance between $x$ and $y$, the algorithm keeps adding $(x_i-y_i)^2$ for $i=1,2,\ldots,\ell$. When the sum exceeds $\delta^2$, this pair is immediately dropped (without completing the rest of the distance computation). This technique is known as {\em early abandoning}.

The second idea in MK uses the triangular inequality in a novel way.
Let $x$ and $y$ be any two $\ell$-mers.  At any stage in the algorithm, we have an upper bound $\delta$ on the distance between the closest pair of $\ell$-mers. If $d(x,y)$ can be inferred to be greater than $\delta$, then we can drop the pair $(x,y)$ from future consideration (since this pair cannot be the closest). Ideally, we would like to calculate $d(x,y)$ exactly for every pair of $\ell$-mers $x$ and $y$. But this will take too much time. MK circumvents this problem by {\bf estimating} the distance between $x$ and $y$ via the triangular inequality. In particular, a random reference $\ell$-mer  $r$ is chosen and the distance between each $\ell$-mer and $r$ is computed.  The $\ell$-mers are kept in an ascending order of their distances to $r$. From thereon, $d(r,y)-d(r,x)$ is used as a lower bound on $d(x,y)$. If this lower bound is $>\delta$, then $(x,y)$ is dropped from future consideration.

The above algorithm is generalized to employ multiple reference $\ell$-mers. The use of multiple references speeds up the algorithm.

\subsection{An Analysis of the MK Algorithm and Our New Idea}
In this section we provide an (informal) analysis of the MK algorithm to explain why the algorithm has a very good performance. Specifically, if we choose multiple random reference points, the algorithm achieves a much better run time than having a single reference point. We explain why this is the case.

\begin{figure}[h]
\includegraphics[scale=.50]{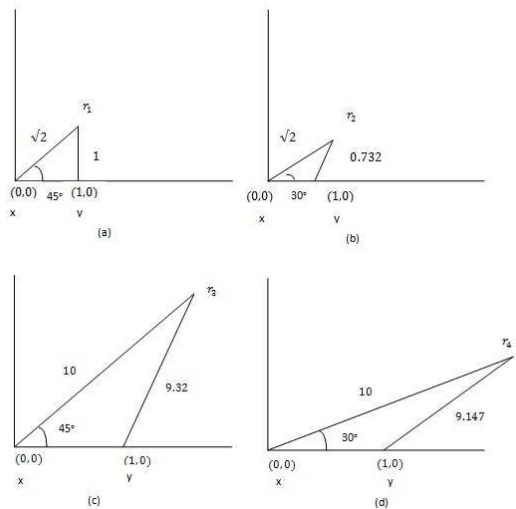}
\caption{Reference Points}\label{figure1}
\label{fig:1}       
\end{figure}

For ease of understanding consider the 2D Euclidean space. The analysis can be extended to points in $\Re^\ell$. For any two $\ell$-mers $x$ and $y$, the closer $d(r,y)-d(r,x)$ is to $d(x,y)$, the better will be our estimate and hence the better will be our chance of dropping $(x,y)$ (if $(x,y)$ is not the closest pair). It turns out that the quality of the lower bound $d(r,y)-d(r,x)$ is decided by two factors: 1) the angle $\angle rxy$ and 2) $d(r,x)$. We illustrate this with an example. Let $x=(0,0)$ and $y=(1,0)$. Consider a reference point $r_1=(1,1)$ (Figure~\ref{figure1}(a)). Note that $r_1$ is at a distance of $\sqrt 2$ from $x$. In this case $d(r_1,x)-d(r_1,y)=0.414$. Also, $\angle r_1xy=45^\circ$. Let $r_2$ be the point we get by keeping the distance between the reference point and $x$ the same, but changing this angle to $30^\circ$ (Figure~\ref{figure1}(b)). In this case,  $d(r_2,x)-d(r_2,y)$ improves to $0.6722$. As another example, if the reference point $r$ lies on the perpendicular bisector of $x$ and $y$, then $d(r,x)-d(r,y)=0.$

For any two input points $x$ and $y$, if we pick multiple reference points randomly, then we would expect that at least one of these reference points $r$ will be such that the angle $\angle rxy$ will be such that $d(r,x)-d(r,y)$ will be `large'. In contrast, if we have only one reference point, for some pairs of points the corresponding angles may be `good', but for a good percentage of the pairs, the angles may not be `good'. (A reference point $r$ is `'good' if $d(r,x)-d(r,y)$ is close to $d(x,y)$).

The effect of $d(r,x)$ on $d(r,x)-d(r,y)$ can be seen with the same examples. Consider the example of Figure~\ref{figure1}(a). Assume that we keep the angle the same but increase $d(r_1,x)$ to $10$ and get the reference point $r_3$ (Figure~\ref{figure1}(c)). In this case, $d(r_3,x)-d(r_3,y)$ improves to $0.6803$. Also, in the example of Figure~\ref{figure1}(b), say we keep the angle the same but increase $d(r_2,x)$ to $10$ and get the reference point $r_4$. In this case, $d(r_4,x)-d(r_4,y)$ improves to $0.8526$. Of course, if the reference point $r$ lies on the perpendicular between $x$ and $y$, then however large $d(r,x)$ could be, $d(r,x)-d(r,y)$ will continue to be zero! However, the probability of this happening is low. For a given angle $\theta$, we can compute the limit of $d(r,x)-d(r,y)$ as $d(r,x)$ tends to $\infty$. For instance when the angle is $45^\circ$ (Figure~\ref{figure1}(a)), this limit is $\frac{1}{\sqrt 2}\approx 0.707$.

\subsection{Our algorithm}
Our proposed new algorithm indeed exploits the relationship between $d(r,x)$ and $d(r,x)-d(r,y)$. In particular, we pick a collection $C$ of random reference points and project each of these points out by multiplying each coordinate value of each point by a factor of $f$. For example, $f$ could be $10$. The rest of the algorithm is the same as MK.

A pseudocode for our algorithm, called {\em Motif discovery with Projected Reference points (MPR)},  is given below.

\noindent {\bf Algorithm} {\sf MPR}\\
{\em Input:} $T=a_1,a_2,\ldots,a_n$ and an integer $\ell$, where each $a_i$ is a real number (for $1\leq i\leq n$). Input are also $q$ and $f$, where $q$ is the number of references and $f$ is the projection factor.\\
{\em Output:} The two closest $\ell$-mers of $T$.\\

\vspace{-0.15in}
 \noindent {\bf 1)} Pick $q$ random $\ell$-mers of $T$ as references; Project these
references by multiplying each element in each $\ell$-mer
 by $f$. Let these projected references be $r_1,r_2,\ldots,r_q$.\\
 {\bf 2)} Compute the distance between every $\ell$-mer of $T$ and
 every projected reference $\ell$-mer.\\
{\bf 3)} Sort the $\ell$-mers of $T$ with respect to their distances
 to $r_1$. Let the sorted $\ell$-mers be
$p_1,p_2,\ldots,p_{n-\ell+1}$.\\
{\bf 4)} Let $\delta=\infty$; Let $answer=(0,0)$;\\
{\bf 5) for} $i:=1$ {\bf to} $(n-\ell+1)$ {\bf do}\\
\hspace*{0.4in} {\bf for} $j:=(i+1)$ {\bf to} $(n-\ell+1)$ {\bf do}\\
\hspace*{0.7in} $failure:=false$;\\
\hspace*{0.7in} {\bf for} $k:=1$ {\bf to} $q$ {\bf do}\\
\hspace*{0.9in} {\bf if} $d(r_k,p_j)-d(r_k,p_i)>\delta$ {\bf then}\\
\hspace*{1.1in} $failure:=true$; {\bf exit};\\
\hspace*{0.7in} {\bf if} $failure$ {\bf then exit else}\\
\hspace*{0.9in} Compute $d(p_i,p_j)$; \\
\hspace*{0.9in} {\bf if} $d(p_i,p_j)<\delta$ {\bf then}\\
\hspace*{1.1in} $\delta:=d(p_i,p_j)$; $answer=(i,j)$;\\
{\bf 6) Output} $(i,j)$.

\noindent{\bf Observation:} Please note that even though the above algorithm has been presented for solving the TSMM problem, it is straight forward to extend it to solve CPP in $\Re^\ell$.

\subsection{An analysis of our algorithm}
In this section we show why our idea of projecting reference points improves the performance of the algorithm. Let the input points be from $\Re^d$ for some integer $d$. Consider any two input points $A$ and $B$. Let $R$ be any reference point. Note that any three points are coplanar. Consider any hyperplane $\cal H$ containing $A,B,$ and $R$. If we multiply every coordinate of $R$ by the same number, then the resultant point will also lie in $\cal H$. This is because the equation defining $\cal H$ will be of the form $a_1x_1+a_2x_2+\cdots+a_dx_d=0$. Thus, in order to see how $d(R,A)-d(R,B)$ changes with a scaling of $R$, it suffices to consider the case that these three points are in 2D.

\begin{figure}[h]
\includegraphics[scale=.50]{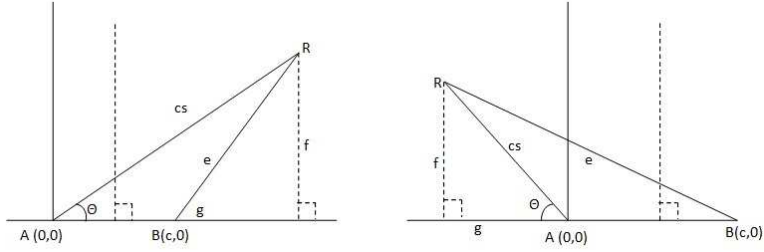}
\caption{The effect of scaling on reference points}\label{fig2}      
\end{figure}

Without loss of generality let $A$ be $(0,0)$ and $B$ be $(c,0)$, for some real number $c$. There are two cases to consider for the position of $R$ relative to $A$ and $B$: 1) $R$ is to the right of the perpendicular bisector of $A$ and $B$; 2) $R$ is to the left of the perpendicular bisector between $A$ and $B$. These two cases are illustrated in Figures~\ref{fig2}(a) and \ref{fig2}(b), respectively. Note that when $R$ lies on the perpendicular bisector of $A$ and $B$, $d(R,A)-d(R,B)$ will be zero.

In case 1, let $d(A,R)$ be $cs$, $s$ being a scaling factor. $f=cs\sin\theta$, $g=cs\cos\theta-c$, and $e=\sqrt{f^2+g^2}$. As a result, $e=\sqrt{c^2s^2+c^2-2c^2s\cos\theta}$ $=cs\sqrt{1+\frac{1}{s^2}-\frac{2\cos\theta}{s}}$. Using the fact that $(1-u)^n\approx 1-nu$ (when $nu<<1$), $e\approx cs+\frac{c}{2s}-c\cos\theta$. Thus, $d(A,R)-d(R,B)\approx c\cos\theta-\frac{c}{2s}$. Clearly, when $c$ and $\theta$ are the same, the value of $d(R,A)-d(R,B)$ increases when $s$ increases.

In case 2, let $d(A,R)$ be $cs$, for a scaling factor of $s$. Clearly, $f=cs\sin\theta$, $g=cs\cos\theta$. Thus, $e=\sqrt{f^2+(g+c)^2}$. Also, $e=\sqrt{c^2s^2+c^2+2c^2s\cos\theta}=cs\sqrt{1+\frac{1}{s^2}+\frac{2\cos\theta}{s}}$. Using the approximation mentioned in case 1, we see that $e\approx cs+\frac{c}{2s}+c\cos\theta$. Therefore, $d(R,B)-d(R,A)\approx \frac{c}{2s}+c\cos\theta$. In this case, when $c$ and $\theta$ are the same, the value of $d(R,B)-d(R,A)$ increases when $s$ decreases.

But for a given reference point, and two input points $A$ and $B$, we do not know which of the two cases will hold. But we can expect that half of the randomly chosen reference points will fall under case 1 and the other half will be expected to fall under case 2. If we only employ a scaling factor $s$ that is greater than one, then, for an expected half of the reference points we expect to see an improvement in the estimate of a lower bound for $d(A,B)$. This explains why our algorithm performs better than MK.

The above analysis can also be used to better understand the MK algorithm.

\subsection{Fixed radius nearest neighbors problem (FRNNP)}
In this problem we are given $n$ points $a_1,a_2,\ldots,a_n$ in $\Re^\ell$ and a radius $R$ (which is a real number) and the problem is to identify the $R$-neighborhood of each input point. If $p$ is an input point, its $R$-neighborhood is defined to be the set of all input points that are within a distance of $R$ from $p$. FRNNP has numerous applications. One of the applications of vital importance is that of molecular simulations.

We can modify {\sf MPR} to solve this problem as well. The modified version is given below. Let $N(i)$ denote the $R$-neighborhood of $a_i,1\leq i\leq n$.

\noindent {\bf Algorithm} {\sf MPR-FRNNS}\\
{\em Input:} $T=a_1,a_2,\ldots,a_n$ and $R$, where each $a_i$ is a point in $\Re^\ell$ (for $1\leq i\leq n$) and $R$ is a real number. Input are also $q$ and $f$, where $q$ is the number of references and $f$ is the projection factor.\\
{\em Output:} $N(i)$, for $1\leq i\leq n$.\\

\vspace{-0.15in}
\noindent {\bf 1)} Pick $q$ random points of $T$ as references; Project these \\
references by multiplying each coordinate of each reference\\
point by $f$. Let these projected references be $r_1,r_2,\ldots,r_q$.\\
 {\bf 2)} Compute the distance between every point of $T$ and \\
 every projected reference point.\\
{\bf 3)} Sort the points of $T$ with respect to their distances\\
 to $r_1$. Let the sorted points be
$p_1,p_2,\ldots,p_n$.\\
{\bf 4) for} $i:=1$ {\bf to} $n$ {\bf do} $N(i):=\emptyset$;\\
{\bf 5) for} $i:=1$ {\bf to} $n$ {\bf do}\\
\hspace*{0.4in} {\bf for} $j:=(i+1)$ {\bf to} $n$ {\bf do}\\
\hspace*{0.7in} $failure:=false$;\\
\hspace*{0.7in} {\bf for} $k:=1$ {\bf to} $q$ {\bf do}\\
\hspace*{0.9in} {\bf if} $d(r_k,p_j)-d(r_k,p_i)>R$ {\bf then}\\
\hspace*{1.1in} $failure:=true$; {\bf exit};\\
\hspace*{0.7in} {\bf if} $failure$ {\bf then exit else}\\
\hspace*{0.9in} Compute $d(p_i,p_j)$; \\
\hspace*{0.9in} {\bf if} $d(p_i,p_j)\leq R$ {\bf then} add $j$ to $N(i)$ and \\
\hspace*{0.9in} add $i$ to $N(j)$;\\
{\bf 6) Output} $N(i),$ for $1\leq i\leq n$.

\subsection{An experimental comparison of MK and MPR}
A typical algorithm in the literature for CPP has two phases. In the first phase pairs of points that cannot possibly be the closest are eliminated. In the second phase distance is computed between every pair of points that survive the first phase. The time spent in the first phase is typically very small and hence is negligible (compared to the time spent in the second phase). Also, the time needed to process the pairs in the second phase is linear in the number of surviving pairs. As a result, it suffices to report the number of pairs (to be processed in the second phase) as a measure of performance (see e.g., \cite{PBK11}). In this paper also we use this measure of performance throughout.

We have experimentally compared the performance of MK and MPR on different data sets. The machine we have used has an
Intel(R) Core(TM) i7-2640M 2.8 GHz CPU with 8GB RAM running Windows 7 (64 bit). The same machine has been used for all the experiments reported in this paper.

As mentioned in \cite{AEQSB09}, random walk data set is the most difficult case for time series mining algorithms since the probability of the existence of very close motif pairs is very low.
We have also used the same data for our comparison.
In particular, we have used 10 different random walk data sets of sizes ranging from 10K to 100K. We have also varied the motif length to see how the performances change. Our algorithm performs better than MK for higher motif lengths. Both the algorithms have been run 10 times and the averages computed. We do not perform any comparison with the brute force  method as that has already been done in \cite{AEQSB09}.\newline

\begin{table}[ht]
\centering  
\scalebox{0.6}{
\begin{tabular}{|c|c|c|c|c|} 
\hline
& & & &\\                        
Dataset Size & Pairs (MK) & Runtime in sec(MK) & Pairs (MPR) & Runtime in sec(MPR)\\ [0.2ex] 
\hline                  
& & & &\\
$10,000$ & $1.0078\times10^7$ & $18.5$ & $0.7634\times10^7$ & $10.1$\\ [0.2ex] 
\hline                  
& & & &\\
$20,000$ & $1.1858\times10^7$ & $60.7$ & $0.7650\times10^7$ & $40.8$\\ [0.2ex] 
\hline                  
& & & &\\
$30,000$ & $2.0001\times10^7$ & $60.9$ & $1.3793\times10^7$ & $42.3$\\ 
\hline                  
& & & &\\
$40,000$ & $3.1785\times10^7$ & $62.3$ & $2.1476\times10^7$ & $43.5$\\ [1ex]      
\hline                  
& & & &\\
$50,000$ & $4.8031\times10^7$ & $101.5$ & $3.2897\times10^7$ & $51.8$\\ [1ex]      
\hline                  
& & & &\\
$60,000$ & $5.1409\times10^7$ & $102.1$ & $3.5454\times10^7$ & $52.3$\\ [1ex]      
\hline                  
& & & &\\
$70,000$ & $1.4083\times10^8$ & $288.9$ & $0.9388\times10^8$ & $167.8$\\ [1ex]      
\hline
& & & &\\
$80,000$ & $1.7896\times10^8$ & $340.9$ & $1.1930\times10^8$ & $232.8$\\ [1ex]      
\hline
& & & &\\
$90,000$ & $2.1198\times10^8$ & $550.3$ & $1.4519\times10^8$ & $341.0$\\ [1ex]      
\hline
& & & &\\
$100,000$ & $3.1949\times10^8$ & $772.5$ & $2.1587\times10^8$ & $520.9$\\ [1ex]      
\hline 
\end{tabular}}
\vspace*{0.2in}
\caption{Number of Pairs and Runtime comparison: Euclidean case}\label{table1} 
\end{table}

\begin{figure}[h]
\includegraphics[scale=.50]{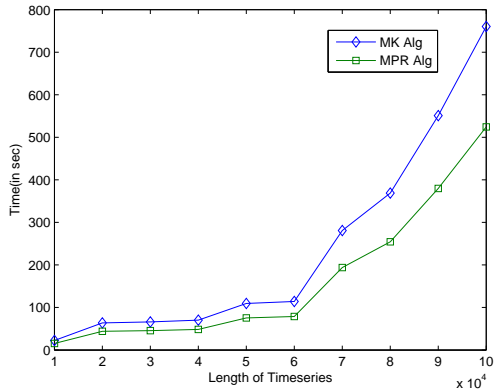}
\caption{Runtime Comparison between MK and MPR: Euclidean case}\label{figure2}
\end{figure}

In Table~\ref{table1} we show the number of pairs processed (in the second phase) in MK and MPR.  The size (i.e., the length)
 of the time series data varies from 10K to 100K, the motif length being 1024. The following parameter values have been used: $q = 10$ and $f=10$.
 From this table we see that MK processes around 1.5 times the number of pairs processed by MPR.
Figure~\ref{figure2} presents a graphical plot of the runtime requirements of MK and MPR algorithms. This figure shows that the run time of MK is around 1.5 times the run time of MPR. This improvement is quite significant for the following reason: There are two ideas used in MK, namely, early abandoning and the use of random reference points. As the authors point out in \cite{AEQSB09}, the difference between using early abandoning alone and both the ideas is small, especially on random walk data sets. However, MK performs much better than using early abandoning alone on real datasets.

\begin{figure}[h]
\includegraphics[scale=.50]{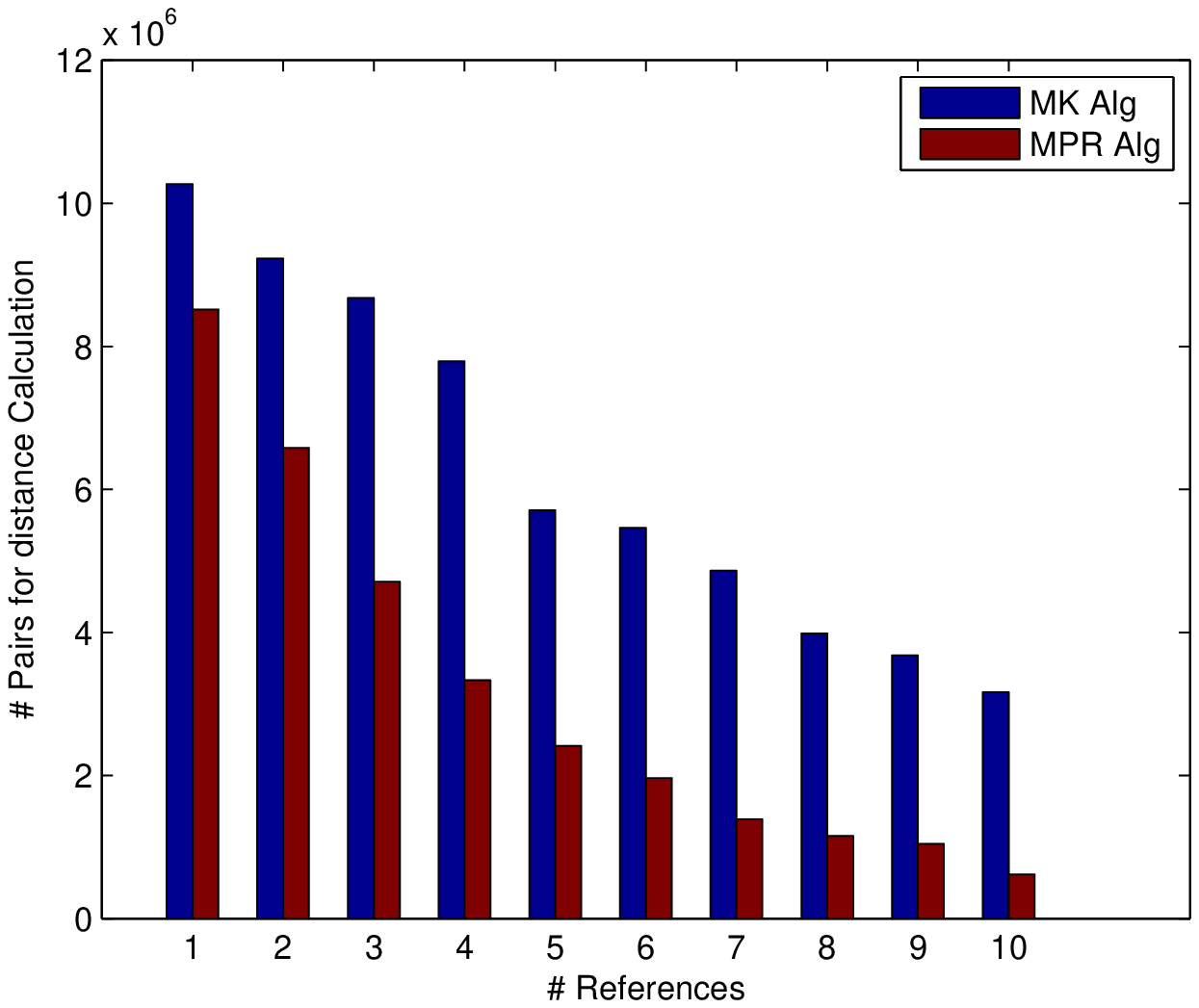}
\caption{Number of pairs as a function of the number of reference points: Euclidean case}\label{figure3}
\end{figure}

Table~\ref{table2} and Figure~\ref{figure3} show how the number of pairs  reduces with an increase in the number $q$ of the reference points for MK and MPR algorithms. From these, we note that as $q$ increases, the difference between MK and MPR widens.

\begin{table}[ht]
\centering  
\scalebox{0.6}{
\begin{tabular}{|c|c|c|c|c|} 
\hline
& & & &\\                        
No of Ref & No of Pairs(MK) & Runtime(MK) & No of Pairs(MPR) & Runtime(MPR) \\ [0.2ex] 
\hline                  
& & & &\\
$1$ & $1.7157462\times10^7$ & 18.93 sec & $1.2112471\times10^7$ & 13.17 sec\\ [0.2ex] 
\hline                  
& & & &\\
$2$ & $0.7806227\times10^7$ & 10.43 sec & $0.3773235\times10^7$ & 5.03 sec\\ [0.2ex] 
\hline                  
& & & &\\
$3$ & $0.3500211\times10^7$ & 5.74 sec & $0.1534813\times10^7$ & 3.15 sec\\ 
\hline                  
& & & &\\
$4$ & $0.1769827\times10^7$ & 4.29 sec & $0.0896315\times10^7$ & 2.23 sec\\ [1ex]      
\hline                  
& & & &\\
$5$ & $0.1255878\times10^7$ & 3.15 sec & $0.0458978\times10^7$ & 2.16 sec\\ [1ex]      
\hline                  
& & & &\\
$6$ & $0.0885219\times10^7$ & 2.88 sec & $0.0314989\times10^7$ & 1.90 sec\\ [1ex]      
\hline                  
& & & &\\
$7$ & $0.0725813\times10^7$ & 2.73 sec & $0.0273630\times10^7$ & 1.73 sec\\ [1ex]      
\hline
& & & &\\
$8$ & $0.0547364\times10^7$ & 2.69 sec & $0.0216532\times10^7$ & 1.68 sec\\ [1ex]      
\hline
& & & &\\
$9$ & $0.0460080\times10^7$ & 2.45 sec & $0.0179179\times10^7$ & 1.60 sec\\ [1ex]      
\hline
& & & &\\
$10$ & $0.0410578\times10^7$ & 2.46 sec & $0.0178203\times10^7$ & 1.51 sec\\ [1ex]      
\hline 
& & & &\\
$20$ & $0.0297705\times10^7$ & 2.37 sec & $0.0173253\times10^7$ & 1.32 sec\\ [0.2ex] 
\hline                  
& & & &\\
$30$ & $0.0353774\times10^7$ & 2.06 sec & $0.0310319\times10^7$ & 1.18 sec\\ [0.2ex] 
\hline                  
& & & &\\
$40$ & $0.0432340\times10^7$ & 1.99 sec & $0.0405625\times10^7$ & 1.14 sec\\ 
\hline                  
& & & &\\
$50$ & $0.0523544\times10^7$ & 1.96 sec & $0.0502900\times10^7$ & 1.09 sec\\ [1ex]      
\hline                  
& & & &\\
$60$ & $0.0615568\times10^7$ & 2.00 sec & $0.0598589\times10^7$ & 1.12 sec\\ [1ex]      
\hline                  
& & & &\\
$70$ & $0.0711500\times10^7$ & 2.01 sec & $0.0696858\times10^7$ & 1.12 sec\\ [1ex]      
\hline                  
& & & &\\
$80$ & $0.0806084\times10^7$ & 2.12 sec & $0.0794800\times10^7$ & 1.12 sec\\ [1ex]      
\hline
& & & &\\
$90$ & $0.0903655\times10^7$ & 2.17 sec & $0.0892789\times10^7$ & 1.19 sec\\ [1ex]      
\hline
& & & &\\
$100$ & $0.1001488\times10^7$ & 2.19 sec & $0.0991276\times10^7$ & 1.19 sec\\ [1ex]      
\hline
\end{tabular}}
\vspace{0.2in}
\caption{Number of pairs as a function of \protect\linebreak the number of references: Euclidean case}\label{table2} 
\end{table}

The bar graph in Figure~\ref{figure3} pictorially represents this comparison. The blue and red bars represent the number of pairs processed by MK and MPR, respectively.

We have run both MK and MPR algorithms on some of the real data sets from \url{http://www.cs.ucr.edu/~mueen/MK/}. Table~\ref{table3} shows the performance of MK and MPR algorithms. On dataset 1, MPR is around 2 times faster than MK and on dataset 2, MPR is around 1.5 times faster than MK. In this Table, `-' indicates that the algorithm did not stop within 40 minutes.

\begin{table}[ht]
\centering  
\scalebox{0.6}{
\begin{tabular}{|c|c|c|c|c|c|} 
\hline
& & & & &\\                        
Dataset & Length & No. of Pairs (MK) & Runtime(MK) & No. of Pairs (MPR) & Runtime(MPR)\\ [0.2ex] 
\hline                  
& & & & &\\
1 & 33,021 & $1.2719\times10^7$ & 43.944 sec & $0.6112\times10^7$ & 25.214 sec\\ [0.2ex] 
\hline                  
& & & & &\\
2 & 18,667 & $2.1927\times10^8$ & 485.591 sec & $1.4570\times10^8$ & 355.531 sec\\ 
\hline                  
& & & & &\\
3 & 78,254 & $19.0220\times10^8$ & 6018.072 sec & $1.5027\times10^9$ & 3857.739 sec\\ [1ex]      
\hline                  
\end{tabular}}
\vspace{0.2in}
\caption{Comparison with Real Data sets. \protect\newline Data set 1: $Repeated Insect Behavior Dataset 1$; \protect\newline Data set 2: $Repeated Insect Behavior Dataset 2$; \protect\newline Data set 3: $Repeated Insect Behavior Dataset 3$}\label{table3} 
\end{table}
\section{The Case of Character Strings}\label{sec4}
In this section we consider the CPP when the space is one of character strings and the metric is Hamming distance. An algorithm for this version of CPP has numerous applications including approximate TSMM (see e.g., \cite{BES03}). When the alphabet is $\{0,1\}$, the light bulb algorithm of \cite{RSJ89} can be used to solve this problem. In this section we show how to modify the light bulb algorithm for the case of generic alphabets. Before presenting details on the modification, we provide a brief summary of the light bulb problem.

\subsection{The light bulb problem}
The light bulb problem is that of identifying the most correlated pair of bulbs from out of $n$ given bulbs \newline $b_1,b_2,\ldots,b_n$. This problem is solved by observing the state of each bulb in $t$ discrete time steps (for some relevant value of $t$). The states of bulb $i$ in these $t$ time steps can be represented as a vector $\hat{b}_i=(b_1^i,b_2^i,\ldots,b_t^i)$ (for $1\leq i\leq n$). We can think of $\hat{b}_i$ as a sample from a probability distribution that governs the state of bulb $i$. It can be shown that if $t$ is sufficiently large (e.g., $\Omega(\log n)$), then the pair of bulbs that is the most correlated in the samples is also the most correlated pair with high probability. Thus the light bulb problem can be stated as follows: We are given $n$ Boolean vectors $\hat{b}_1,\hat{b}_2,\ldots,\hat{b}_n$. The problem is to find the pair of vectors that are the most similar (i.e., the Hamming distance between them is the smallest). From hereon, we will use this formulation of the problem.

   Note that, given two vectors, we can find the Hamming distance between them in $O(t)$ time. A straight forward algorithm to identify the most correlated pair of bulbs takes $O(n^2t)$ time. This algorithm computes the Hamming distance between every pair of bulbs. The algorithm of \cite{RSJ89} takes subquadratic time. In particular, the expected run time of this algorithm is \newline $O\left (n^{1+\frac{\log p_1}{\log p_2}}\log^2n\right )$ assuming that $t=O(\log n)$. Here, $p_1$ is the correlation between the most correlated pair of bulbs and $p_2$ is the correlation between the second most correlated pair of bulbs. Note that if the correlation between two bulbs $i$ and $j$ is $p_{ij}$ then the expected Hamming distance between $\hat{b}_i$ and $\hat{b}_j$ is $t(1-p_{ij})$. Equivalently, the similarity (i.e., the number of matches) between $\hat{b}_i$ and $\hat{b}_j$ is $tp_{ij}$.

\subsection{The light bulb algorithm}
Consider a matrix $M$ of size $n\times t$, such that the $i$th row of $M$ is $\hat{b}_i$, for $1\leq i\leq n$. The algorithm of \cite{RSJ89} iteratively collects pairs of bulbs that are candidates to be the most correlated. Once it collects enough pairs, it computes the distance between each pair in this collection and outputs the closest. There are $O\left (n^{\frac{\log p_1}{\log p_2}}\log n\right )$ iterations in the algorithm and in each iteration, some candidate pairs are generated and added to the collection $C$. In any iteration, the algorithm picks $c\log n$ columns of $M$ at random (for some constant $c$). The rows are sorted based on the characters in the randomly chosen columns. As a result of this sorting, the bulbs get partitioned into buckets such that all the bulbs with equal values (in the $c\log n$ random columns) fall into the same bucket. A pair of bulbs $(a,b)$ will get added to $C$ in any iteration if they fall into the same bucket in this iteration.

The authors of \cite{RSJ89} show that after $O\left (n^{\frac{\log p_1}{\log p_2}}\log n\right )$ iterations, $C$ will have the most correlated pair of bulbs with high probability (i.e., with a probability of $1-n^{-\Omega(1)}$). The above algorithm has been proposed for the case of binary strings. We can modify this algorithm to handle the case of an arbitrary (finite) alphabet and get the following theorem.

\begin{theorem}\label{theorem0}
Let $M$ be a matrix of size $n\times t$. Each entry in this matrix is an element from some set $\Sigma$ of cardinality $\sigma$. We can find the most correlated pair of columns of $M$ in an expected $O\left (n^{1+\frac{\log p_1}{\log p_2}}\frac{\log^2 n\log\sigma}{w}\right )$ time where $p_1$ is the correlation between the most correlated pair of columns, $p_2$ is the correlation between the second most correlated pair of columns, and $w$ is the word length of the machine.  This expected run time will be $O\left (n^{1+\frac{\log p_1}{\log p_2}}\log^2 n\right )$ if we use a general sorting algorithm. (Here correlation is based on Hamming distance. For example, $p_1$ is the largest fraction of rows in which any two columns agree).
\end{theorem}


\noindent{\bf Proof:} The only difference in the algorithm is that instead of sorting binary strings we will have to sort strings from an arbitrary alphabet. Without loss of generality, let $\Sigma=\{0,1,2,\ldots,\sigma-1\}$ be the alphabet under concern. In the original algorithm, one has to sort $n$ $(c\log n)$-bit integers in every iteration. For a generic alphabet, we have to sort $n$ $(c\log n\log \sigma)$-bit integers. If one uses any comparison based sorting algorithm, this sorting takes $O(n\log n)$ time. If we use an integer sorting algorithm, this sorting can be done in $O\left (\frac{cn\log n\log \sigma}{w}\right )$ time where $w$ is the word length of the machine. This is the time spent in each iteration of the algorithm. Therefore, the total expected run time is $O\left (n^{1+\frac{\log p_1}{\log p_2}}\frac{\log^2 n\log\sigma}{w}\right )$. $\Box$

Let this modified version of the light bulb algorithm be called MLBA.

\subsection{An experimental comparison}
The algorithm of \cite{BES03} for approximate TSMM partitions the input time series data $T$ based on a window of size $w$ (for an appropriate value of $w$), computes the mean of every window, and discretizes the mean into four possible values. As a result, the time series data is transformed into a string $T'$ of characters form the alphabet $\{1,2,3,4\}$. It then uses any $(\ell,d)$-motif finding algorithm to find the motifs in $T'$. However, all the exact algorithms for finding $(\ell,d)$-motifs take time that is exponential on $\ell$ and $d$. Note that the last step of finding $(\ell,d)$ motifs can be replaced with a problem of finding time series motifs in $T'$ which is nothing but CPP in the domain of strings of characters, the motif length being $\ell$.

One could employ MK to solve CPP in the domain of character strings. The only difference is that we have to replace Euclidean distance with Hamming distance. We have implemented this algorithm. It turns out that MK does not perform well for the case of Hamming distance. To be fair, the authors of MK have not tested MK for this case. We have compared MK with MLBA and the results are shown in Table~\ref{table4}. As this Table reveals, MLBA is around 200 times faster than MK. It is also clear that if we employ MLBA in place of $(l,d)$-motif finding algorithms, the performance of the approximate TSMM algorithm given in \cite{BES03} will improve significantly.

\begin{table}[ht]
\centering  
\scalebox{0.6}{
\begin{tabular}{|c|c|c|c|c|} 
\hline
& & & &\\                        
Dataset Size & Pairs(MK) & Runtime in sec(MK) & Pairs(MLBA) & Runtime in sec(MLBA)\\ [0.1ex] 
\hline                  
& & & &\\
$1,000$ & 176,799 & $0.0067$ & 2,012 & $0.0002$\\ [0.1ex] 
\hline                  
& & & &\\
$2,000$ & 1,981,048 & $10.1$ & 7,849 & $0.0003$\\ [0.1ex] 
\hline                  
& & & &\\
$3,000$ & 5,783,132 & $16.5$ & 17,795 & $0.0005$\\ 
\hline                  
& & & &\\
$4,000$ & 11,586,390 & $39.8$ & 31,337 & $0.0089$\\ [1ex]      
\hline                  
& & & &\\
$5,000$ & 9,693,660 & $55.3$ & 48,867 & $1.3$\\ [1ex]      
\hline                  
& & & &\\
$6,000$ & 14,594,238 & $86.7$ & 71,164 & $1.5$\\ [1ex]      
\hline                  
& & & &\\
$7,000$ & 20,494,808 & $124.3$ & 96,539 & $1.7$\\ [1ex]      
\hline
& & & &\\
$8,000$ & 27,395,331 & $161.2$ & 125,343 & $1.9$\\ [1ex]      
\hline
& & & &\\
$9,000$ & 35,295,659 & $217.8$ & 317,220 & $2.1$\\ [1ex]      
\hline
& & & &\\
$10,000$ & 44,195,948 & $276.9$ & 602,326 & $2.1$\\ [1ex]      
\hline 
\end{tabular}}
\vspace{0.2in}
\caption{Number of Pairs and Runtime comparison on \protect\linebreak strings and Hamming distance}\label{table4} 
\end{table}

\begin{figure}[h]
\includegraphics[scale=.50]{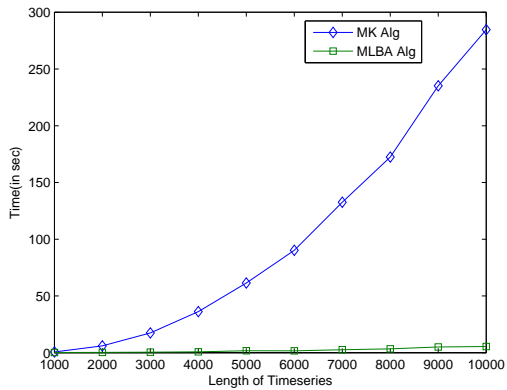}
\caption{Runtime comparison between MK and MLBA \protect\linebreak for character strings}\label{figure4}
\end{figure}

Figure~\ref{figure4} shows a runtime comparison of MK and MLBA for the case of character strings from a finite alphabet.

\section{Identification of the Least Correlated Pair of Strings}\label{sec5}
The light bulb algorithm of \cite{RSJ89} identifies the closest pair of strings, from out of $n$ given binary strings. An interesting question is if we can use the same algorithm to identify the {\bf furthest} pair of strings. This problem has relevance in many problems including the two locus problem in GWAS.
The authors of \cite{PBK11} present an elegant adaptation of the light bulb algorithm to solve this problem when the strings are binary. They also show how to solve this problem for arbitrary alphabets using Locality Sensitive Hashing (LSH) \cite{M02}. They map the input strings into binary strings using LSH. LSH closely preserves similarities with a high probability. In this section we show how to avoid LSH. In particular, we present novel deterministic mappings of the input strings to binary strings such that similarities are preserved deterministically.  Our experimental comparison shows that our algorithm has a  significantly better run time than that of \cite{PBK11}.

\subsection{Some notations}
 Let $m(x,y)$ stand for the number of matches between two strings (of equal length) $x$ and $y$. For instance, if $x=10010$ and $y=00111$, then $m(x,y)=2$ (since they match in positions 2 and 4). Let $X=x_1,x_2,\ldots,x_q$ and $Y=y_1,y_2,\ldots,y_q$ be two sequences of strings (each string having the same length). We define $M(X,Y)$ to be $\sum_{i=1}^qm(x_i,y_i)$.

   Consider the sequences $A_i=a_1^i,a_2^i,\ldots,a_k^i$, for $1\leq i\leq n$, where each $a_j^i$ is $0,1,$ or $2$ (for $1\leq j\leq k$). Note that each $A_i$ is a sequence of strings where each string is of length 1. Let $M(A_i,A_j)=u_{ij}$.

Each $A_i$ can be thought of as a string from the alphabet $\{0,1,2\}$. In the application of GWAS, we can let $A_i$ correspond to the SNP $i$, for $1\leq i\leq n$. Specifically, $a_j^i$ is the value of the $i$th SNP in subject $j$, for $1\leq j\leq k$. If we are interested in finding the two most correlated SNPs, then we can use MLBA to identify this pair (as shown in Section~\ref{sec4}). On the other hand, if our goal is to identify the least correlated pair, then, it is not clear how to do this using MLBA. To solve the two locus GWAS problem, we have to identify not only the most correlated pair of bulbs but also the least correlated pair.

\subsection{Finding the least correlated pair -- the case of zeros and ones}
 The authors of \cite{PBK11} present an elegant solution for this problem when each $A_i$ has only zeros and ones. In this case, each $A_i$ can be thought of as a light bulb. The idea is to construct a matrix $D$ of size $k\times 2n$ where each column of $D$ corresponds to either a bulb or its `complement', Specifically, the first $n$ columns correspond to the bulbs and the next $n$ columns correspond to the complements of the bulbs. In other words, $D[j,i]=a_j^i$, for $1\leq i\leq n, 1\leq j\leq k$ and $D[j,i]=\bar{a}_j^i$ for $1\leq j\leq k,(n+1)\leq i\leq 2n$. Here, if $x$ is any bit, then, $\bar{x}$ denotes its complement. Let $D_1=\{q:1\leq q\leq n\}$ and $D_2=\{q:(n+1)\leq q\leq 2n\}$. The algorithm of \cite{PBK11} for finding the least correlated pair works as follows. Consider all the pairs of columns $(a,b)$ such that $a\in D_1$ and $b\in D_2$. From out of these pairs, identify the pair $(a',b')$ of columns with the maximum number of matches.  If $a'=i$ and $b'=n+j$, then $(i,j)$ is the least correlated pair of bulbs. Finding such a pair $(a',b')$ can be done using the light bulb algorithm of \cite{RSJ89}. The correctness of this algorithm follows from the fact that if the two bulbs $i$ and $j$ have the least number of matches, then, column $i$ and the complement of column $j$ will have the most number of matches.

\subsection{Finding the least correlated pair - the case of zeros, ones, and twos}\label{mapping}
 It is not clear how to extend the above idea when the sequences have three (or more) possible elements. The authors of \cite{PBK11} reduce such general cases to the case of zeros and ones using locality sensitive hashing (LSH).
 The measure of correlation used by \cite{PBK11} is different from what we use in this paper. We define the correlation between two strings $A_i$ and $A_j$ as $p_{ij}=\frac{M(A_i,A_j)}{k}$. In contrast, \cite{PBK11} use Pearson's correlation coefficient.


 In this section we present an elegant algorithm for the problem of identifying the least correlated pair of strings without employing LSH. The idea of \cite{PBK11} is to map input strings into Boolean vectors. If $i$ and $j$ are any two strings, then the sequences $A_i$ and $A_j$ are mapped to Boolean vectors $A_i^\prime$ and $B_i^\prime$ by LSH such that the distance between $A_i$ and $A_j$ will be nearly the same as the distance between $A_i^\prime$ and $A_j^\prime$ with some probability. The larger the length of $A_i^\prime$ is, the better will be the accuracy of LSH in preserving distances.

 Our algorithm also maps each $A_i$ into a Boolean vector $A_i^\prime$ deterministically such that $|A_i^\prime|=3|A_i|$, for $1\leq i\leq n$.

 Consider an alphabet $\Sigma$ with three strings where $\Sigma=\{001,010, 100\}$. Clearly, $m(x,y)=3$ if $x=y$ and $m(x,y)=1$ if $x\neq y$ for any $x,y\in\Sigma$. Also, $m(x,\bar{y})=0$ if $x=y$ and $m(x,\bar{y})=2$ if $x\neq y$. Here $\bar{y}$ stands for the string obtained from $y$ by complementing each bit. For example, if $y=010$ then $\bar{y}=101$.

 Consider the sequences $A_i=a_1^i,a_2^i,\ldots,a_k^i$, for $1\leq i\leq n$, where each $a_j^i$ is $0,1,$ or $2$ (for $1\leq j\leq k$). Note that each $A_i$ is a sequence of strings where each string is of length 1. Let $M(A_i,A_j)=u_{ij}$. Assume now that we encode each $a_j^i$ as follows (for $1\leq i\leq n$ and $1\leq j\leq k$): $0\rightarrow 001; 1\rightarrow 010$: and $2\rightarrow 100$. Let the encoded version of $A_i$ be denoted as $A_i^\prime$ for $1\leq i\leq n$. Note that $|A_i^\prime|=3k$, for any $1\leq i\leq n$.

 It is easy to see that $M(A_i^\prime,A_j^\prime)=3u_{ij}+(k-u_{ij})=k+2u_{ij}$, for any $i$ and $j$ ($1\leq i,j\leq n$). For any $A_i^\prime=a_1^{\prime i},a_2^{\prime i},\ldots, a_{3k}^{\prime i}$, let $\bar{A}_i^\prime=\bar{a}_1^{\prime i},\bar{a}_2^{\prime i},\ldots,\bar{a}_{3k}^{\prime i}$, for $1\leq i\leq n$. Clearly, $M(A_i^\prime,\bar{A}_j^\prime)=2(k-u_{ij}),$ for any $1\leq i,j\leq n$.

Clearly, the following statement is true: If, from out of all the pairs of strings, $(i,j)$ has the largest correlation, i.e., $u_{ij}$ is the largest, then from out of all the Boolean vectors generated, $A_i^\prime$ and $A_j^\prime$ will have the largest correlation. Also, if $u_{ij}$ is the smallest, then, $A_i^\prime$ and $\bar{A}_j^\prime$ will have the largest correlation (from out of the pairs $(A_i^\prime,\bar{A}_j^\prime),i\neq j,1\leq i,j\leq n$).

 We can now form a matrix $D$ of size $3k\times 2n$ where the first $n$ columns correspond to (transformed) strings and the next $n$ columns correspond to complements of (transformed) strings. Let $D_1=\{q:1\leq q\leq n\}$ and $D_2=\{q:(n+1)\leq q\leq 2n\}$.  Consider all the pairs of columns $(a,b)$ such that $a\in D_1$ and $b\in D_2$. From out of these pairs, identify the pair $(a',b')$ of columns with the maximum number of matches.  If $a'=i$ and $b'=n+j$, then $(i,j)$ is the least correlated pair of strings. Finding such a pair $(a',b')$ can be done using the light bulb algorithm of \cite{RSJ89}.

 \subsection{Run Time Analysis}\label{mapanalysis}
\begin{theorem}\label{theorem1}
 Given $n$ strings, we can find the closest pair of strings in an expected time of $O\left (n^{1+\frac{\log p_1}{\log p_2}}\log^2n\right )$, where $p_1$ and $p_2$ are the largest and the second largest correlation values, respectively. Also, we can find the least correlated pair of strings in an expected time of \protect\linebreak $O\left (n^{1+\frac{\log ((2/3)(1-c_1))}{\log ((2/3)(1-c_2))}}\log^2n\right )$, where $c_1$ and $c_2$ are the smallest and the next smallest correlation values, respectively.
 \end{theorem}

\noindent{\bf Proof:}
 When we transform input strings to binary sequences, the ordering of pairs is preserved in terms of correlations as we have shown before. Let $p_1$ be the correlation of the largest correlated pair and $p_2$ be the correlation of the second largest correlated pair. How do these values change in the transformed domain? If $p_1^\prime$ and $p_2^\prime$ are the transformed values of these correlations, respectively, it can be seen that $p_1^\prime=\frac{1}{3}+\frac{2}{3}p_1$ and $p_2^\prime=\frac{1}{3}+\frac{2}{3}p_2$.

 If $c_1$ and $c_2$ are the correlations of the smallest and the second smallest correlated pairs, respectively, and if $c_1^\prime$ and $c_2^\prime$ are the transformed values of these, respectively, then we can see that: $c_1^\prime=\frac{2}{3}(1-c_1)$ and $c_2^\prime=\frac{2}{3}(1-c_2)$. To find the largest correlated pair, we can use MLBA (Theorem~\ref{theorem0}). We use the mapping only to find the least correlated pair. $\Box$

 \subsection{The case of a general alphabet}
 We have thus far considered the case where the alphabet is $\{0,1,2\}$. We can extend the mapping to a general alphabet and get the following theorem.

\begin{theorem}
 Given $n$ strings, we can find the largest correlated pair of strings in an expected time of  \newline $O\left (n^{1+\frac{\log p_1}{\log p_2}}\log^2n\right )$, where $p_1$ and $p_2$ are the largest and the second largest correlation values, respectively. Also, we can find the least correlated pair of strings in an expected  time of \protect\linebreak $O\left (n^{1+\frac{\log ((2/\sigma)(1-c_1))}{\log ((2/\sigma)(1-c_2))}}\log^2n\right )$, where $c_1$ and $c_2$ are the smallest and the next smallest correlation values, respectively, and there are $\sigma$ characters in the alphabet.
 \end{theorem}

\noindent{\bf Proof:} Consider sequences from the alphabet $\{0,1,\ldots,\sigma-1\}$. In this case we map each element of this alphabet to a binary string of length $\sigma$ where there is only one 1. Specifically, we use the following mapping: $0\rightarrow 00\cdots001;~1\rightarrow 00\cdots010;$ etc. As before, we don't need any mapping if our goal is to find the largest correlated pair. The mapping is used only to find the least correlated pair. $\Box$

 We can improve the above theorem by employing a random mapping as follows: We will use a binary string of length $\sigma$ to encode each symbol in the alphabet. The encoding for each symbol is obtained by (uniformly) randomly choosing each bit in the string (of length $\sigma$). Let $x$ and $y$ be any two symbols in the alphabet (with $x\neq y$) and let $e_x$ and $e_y$ be their encodings, respectively. Then, clearly, the expected value of $m(e_x,e_y)$ is $\frac{\sigma}{2}$. Also, the expected value of $m(e_x,\bar{e}_y)$ is $\frac{\sigma}{2}$. If $c$ is the correlation between a pair of strings and if $c^\prime$ is the transformed value, then, it follows that the expected value of $c^\prime$ is $ \frac{1}{2}(1-c)$. An application of the Chernoff bounds will readily imply that the value of $c^\prime$ will indeed be very close to this expected value with a probability of $1-\sigma^{-\Omega(1)}$. Therefore, we get:

  \begin{theorem}
 Given $n$ strings,  we can find the least correlated pair of strings in an expected time of \protect\linebreak \newline $O\left (n^{1+\frac{\log ((1/2)(1-c_1))}{\log ((1/2)(1-c_2))}}\log^2n\right )$, where $c_1$ and $c_2$ are the smallest and the next smallest correlation values, respectively, and there are $\sigma$ characters in the alphabet. $\Box$
 \end{theorem}

\section{Two Locus Association Problem}\label{sec6}
The two locus association problem is defined as follows.
Input is a matrix $M$ of size $(m_1+m_2)\times n$ where $m_1 + m_2$ is the number of patients (subjects) each with $n$ SNPs. Here $m_1$ is the number of cases and $m_2$ is the number of controls. There are three possible values for each SNP, namely, $0,1,$ or $2$.  The cases are of phenotype $1$ and the controls are of phenotype $0$. Rows $1$ through $m_1$ of $M$ correspond to cases. Let this submatrix be called $A$. Rows $m_1+1$ through $m_1+m_2$ of $M$ correspond to controls and let this submatrix be called $B$. Each column of $M$ corresponds to an SNP.  The two locus association problem is to identify the pair of SNPs whose statistical correlation with phenotype is maximally different between cases and controls. As mentioned in \cite{PBK11}, the goal is to identify the pair:

$$\underset{i,j}{\arg\max}\ |P_{A}(i,j)-P_{B}(i,j)|.$$

If $Q$ is any matrix, then, $P_Q(i,j)$ stands for the correlation between the columns $i$ and $j$ of $Q$.

The algorithm of \cite{PBK11} exploits the light bulb algorithm of \cite{RSJ89} and locality sensitive hashing (LSH) \cite{M02}. They use LSH to transform matrices $A$ and $B$ to $A'$ and $B'$, respectively. In particular, each column $c_i$ of $A$ is converted to a column $c_i'$ of zeros and ones. The size of $c_i$ is $1\times m_1$ and the size of $c_i'$ is chosen to be $u=\max\{m_1,m_2\}$. The matrix $B$ is also transformed into $B'$ in a similar manner using LSH. Followed by this, the pair of interest is identified.

To be precise, using $A'$ and $B'$, the matrix $D$ is formed where
$$D=\left[ \begin{array}{cc}
A' & A' \\
B' & \bar{B}' \\
\end{array} \right] $$

where $\bar{B}'$ is obtained from $B'$ by complementing every element of $B'$. Note that $D$ is of size $2u\times 2n$. Let $D_1=\{1,2,\ldots,n\}$ and $D_2=\{n+1,n+2,\ldots,2n\}$. Consider all the pairs of columns $(i,j)$ such that $i\in D_1$ and $j\in D_2$. From out of these pairs, identify the pair $(i',j')$ of columns with the maximum number of matches.  If $i'=a$ and $j'=n+b$, then $(a,b)$ is the pair of interest. This pair can be found using the light bulb algorithm of \cite{RSJ89}.

We can use our mapping ideas to get the following theorem.

\begin{theorem}\label{theorem6.1}
We can find the pair $(i,j)$ of SNPs that maximizes $P_A(i,j)-P_B(i,j)$ in an expected time of $O\left (n^{1+\frac{\log ((1/2)+(p_1/3))}{\log ((1/2)+(p_2/3))}}\log^2n\right )$, where $p_1$ and $p_2$ are the smallest and the next smallest values of $P_A(i,j)-P_B(i,j)$, respectively, over all possible pairs $(i,j)$ of SNPs.
 \end{theorem}

\noindent{\bf Proof:} Our algorithm also uses the same method except that instead of using LSH to map $A$ and $B$ to $A'$ and $B'$, respectively, we employ the deterministic mapping we have proposed in Section~\ref{mapping}.

Let $i$ and $j$ be two SNPs (i.e., two columns in $A$ and $B$). Let $P_A(i,j)=p$ and $P_B(i,j)=q$. Now consider columns $i$ and $n+j$ of $D$. What can we say about the correlation of these columns? From the discussion in Section~\ref{mapanalysis}, we realize that this correlation is $\frac{1}{2}+\frac{1}{3}(p-q)$. This also proves the correctness of our algorithm. Let $p_1$ be the maximum value of $P_A(i,j)-P_B(i,j)$ over all possible pairs $(i,j)$ and let $p_2$ be the second largest value. Then, the run time follows from Theorem~\ref{theorem0}. $\Box$

 In a similar manner we can also find the pair that maximizes $P_B(i,j)-P_A(i,j)$ and hence identify the pair that maximizes $|P_A(i,j)-P_B(i,j)|$.
\subsection{An experimental comparison}
The notion of similarity (between two SNPs) used in \cite{PBK11} is Pearson's correlation coefficient. In this paper the similarity we use is based on the Hamming distance.  Specifically, the complement of the Hamming distance is the measure of similarity we employ.
The authors of \cite{PBK11} have tested their algorithms on different data sets (including random data). Since we do not have access to either these data sets or their algorithms, the only comparison we can do was on the random data. As explained in \cite{PBK11}, we have also generated SNPs from binomial distributions. In particular, for each subject, the value of each SNP is chosen uniformly randomly to be either 0 or 1 with equal probability. This dataset is called NOISE Data in \cite{PBK11} (c.f. Table~\ref{table6} in \cite{PBK11}).  Like in \cite{PBK11}, we have also generated data of sizes $10$K, $50$K, and $100$K. For each size we have generated two different data sets and computed the average number of pairs to be processed. We compare these numbers with the ones reported in \cite{PBK11}. As can be seen from Table~\ref{table5}, our algorithm is around 4 times faster than the one in \cite{PBK11}. Note that this a significant improvement since the typical processing times for the two locus problem are quite high. For example, the authors of \cite{PBK11} report that on some of the data sets (with no more than $10^7$ SNPs), the brute force algorithm for the two locus problem took several days on 1000 CPUs! Thus any improvement in the run time could make a noticeable difference.

How does one ensure that the output of an algorithm for the two locus problem is correct? For small data sizes, one could run the exhaustive brute force algorithm to identify the correct pair and use it to verify correctness. In fact when the number of SNPs is either $10$K or $50$K, we first found the correct answer and then used it to measure the run time of our algorithm as follows. We'll run our algorithm one iteration at a time until the correct pair(s) is (are) picked up by our algorithm. At this point we will stop and report the total number of pairs collected. The numbers shown in Table~\ref{table5} have been obtained in this manner. We could not use this method for $n>50$K, since the brute force algorithm was taking too much time.

When $n$ is very large, we {\em inject} pairs with known correlations. As an example, consider the problem of finding the largest correlated pair of columns in a $m\times n$ matrix $A$. Say we generate each column by picking each element to be either 0 or 1 with equal probability. For any two columns, clearly, the expected correlation is $\frac{1}{2}$. We can perform a probabilistic analysis to get a high probability bound on the largest correlation between any two columns. One could also get this estimate empirically. For example, for $n=10,000$, we generated several random data sets and computed the largest correlation in each and calculated an average. The average maximum correlation was $58.3\%$. Let $p$ be this value. To inject a pair with a correlation of $p'$ where $p'$ is $>p$ we generate a column $a$ with all ones and another column $b$ with $p'm$ ones and $m(1-p')$ zeros. We then replace (any) two columns of $A$ with $a$ and $b$. Clearly, the correlation between $a$ and $b$ is $p'$. The expected correlation between $a$ and any other column of $A$ (other than $b$) is $\frac{1}{2}$. Similarly, the expected correlation between $b$ any other column of $A$ (other than $a$) is $\frac{1}{2}$. Thus the pair $(a,b)$ is likely to be the winner with high probability. We stop our algorithm when this pair is picked by our algorithm. We have picked a value for $p'$ that is only slightly larger than $p$ so as to get an accurate estimate on the run time.

We have used a similar technique to inject pairs for the two locus problem as well. In Table~\ref{table6} we show the results for our algorithm. In these cases we have generated the SNPs randomly from a binomial distribution as before. We have employed 200 cases and 200 controls. Clearly, the expected value of $(p-q)$ is zero for this data. If we map this data using our deterministic mapping, then the expected correlation between any two SNPs will be $\frac{1}{2}$. Here again we empirically found that the largest correlation was around 58.3\% (when $n$ was $10$K). Therefore, we  have injected a pair whose correlation was $60$\%. This pair is likely to be the winner. Our algorithm was run until this pair was picked. At this time, the algorithm was stopped. We also checked if the algorithm picked any pair whose correlation was better than that of the injected one and found none. In all the datasets we tried, we were always able to find a pair (other than the injected one) whose correlation was very close to 60\% and hence the numbers shown in Table~\ref{table6} are very close to the case with no injections.

\begin{table}[ht]
\centering  
\scalebox{0.6}{
\begin{tabular}{|c|c|c|c|c|} 
\hline
& & & &\\                        
Dataset Size & Pairs (GWAS) & Pairs (MLBA) & Top 10 & Top 100\\ [0.2ex] 
\hline                  
& & & &\\
$10,000$ & 7,082,458 & 1,904,999 & 0.6 & 0.53\\ [0.2ex] 
\hline                  
& & & &\\
$50,000$ & 13,626,181 & 53,907,259 & 0.4 & 0.34\\ [0.2ex] 
\hline                  
\end{tabular}}
\vspace{0.2in}
\caption{No of pairs comparison on \protect\linebreak NOISE data set}\label{table5} 
\end{table}

\begin{table}[ht]
\centering  
\scalebox{0.6}{
\begin{tabular}{|c|c|} 
\hline
&\\                        
Dataset Size & No. of Pairs (MLBA) \\ [0.2ex] 
\hline                  
&\\
$10,000$ & 560426 \\ [0.2ex] 
\hline                  
& \\
$50,000$ & 706197 \\ [0.2ex] 
\hline                  
\end{tabular}}
\vspace{0.2in}
\caption{No of pairs comparison on \protect\linebreak synthetic data with injected pairs}\label{table6} 
\end{table}

\noindent{\bf Practical Considerations:} Another important question is for how long we should run the program before we can be sure that the correct pair has been obtained (with a high confidence). Please note that we do not know the values of $p_1$ and $p_2$ (c.f. Theorem~\ref{theorem6.1}). Theorem~\ref{theorem6.1} suggests that the run time of our algorithm is $O(n^{1+\alpha})$ for some relevant $\alpha$. We can empirically estimate $\alpha$. The idea is to measure the run time of the algorithm for various values of $n$ (as explained above), the maximum value of $n$ being as much as possible (for the given computing platform and time constraints). Then we could use any numerical procedure to estimate $\alpha$.

\section{Conclusions}\label{sec7}
In this paper we have presented novel algorithms for the closest pair problem (CPP). CPP is a ubiquitous problem that has numerous applications in varied domains. We have offered algorithms for the cases of Euclidean as well as Hamming distances. We have applied our algorithms for two well studied and important problems, namely, time series motif mining and two locus genome wide association study. Our algorithms significantly improve the best-known algorithms for these problems. Specifically, we improve the results presented in many prior papers including \cite{AEQSB09}, \cite{BES03}, and  \cite{PBK11}.

\section{Acknowledgement}
This research has been supported in part by the NIH grant NIH-R01-LM-010101.









\bibliographystyle{elsarticle-num}

\end{document}